\documentclass{PoS}

\def \xmm {{\it XMM--Newton}}

\def \hcm {\hbox {\ifmmode $ atom cm$^{-2}\else atom cm$^{-2}$\fi}}

\title{EXIST perspective for Supergiant Fast X-ray Transients}

\ShortTitle{EXIST perspective for Supergiant Fast X-ray Transients}

\author{\speaker{Lara Sidoli}\\ %
        INAF/IASF Milano Italy\\
        E-mail: \email{sidoli@iasf-milano.inaf.it}}

\author{Vito Sguera \\
        INAF/IASF Bologna Italy\\
        E-mail: \email{sguera@iasfbo.inaf.it}}

\author{Angela Bazzano \\
        INAF/IASF Roma Italy\\
        E-mail: \email{angela.bazzano@iasf-roma.inaf.it}}

\author{Pietro Ubertini \\
        INAF/IASF Roma Italy\\
        E-mail: \email{pietro.ubertini@iasf-roma.inaf.it}}

\abstract{
 Supergiant Fast X--ray Transients (SFXTs) are one of the 
most intriguing (and unexpected) results 
 of the $INTEGRAL$ mission. They are a new class of High Mass X--ray Binaries involving about 20 sources 
 to date, with 8 firmly identified SFXTs and many candidates. 
 They are composed by a massive OB supergiant star
 as companion donor and  a compact object. At least four SFXTs host a neutron star, because 
 X--ray pulsations have been discovered, while for the others 
 a black hole cannot be excluded.  SFXTs display short X--ray outbursts (compared with Be/X-ray 
 transients) characterized by fast flares on brief timescales of hours and large flux variability 
 typically in the range 1,000-100,000. 
 The $INTEGRAL$/IBIS sensitivity allowed to catch only the 
bright flares (peaking at 10$^{36}$--10$^{37}$~erg~s$^{-1}$),
without persistent or quiescent emission.
The investigation of their properties, in particular 
the rapid variability time scales of their flaring activity, will greatly
benefit from observations with the Energetic X--ray Imaging Survey Telescope 
($EXIST$), with the possibility to perform a 
long term and continuous as possible monitoring of the  hard X--ray  sky.}

\FullConference{The Extreme sky: Sampling the Universe above 10 keV \\
		 October 13-17 2009\\
		 Otranto (Lecce) Italy}

\begin{document}

\section{SFXTs properties}

In the last years a new class of fast X--ray transient sources,
composed of an accreting  compact object and an OB supergiant companion, 
has been discovered mainly by the IBIS instrument on board the
$INTEGRAL$ satellite: the so called Supergiant Fast X--ray Transients (SFXTs, Sguera et al. 2005). 
They spend most of the time in a low level X--ray activity with luminosities values 
in the range 10$^{32}$-10$^{34}$ erg s$^{-1}$,  only occasionally undergoing  bright
(L$_x$=10$^{36}$-10$^{37}$ erg s$^{-1}$) and fast flaring activity, part of outbursts with a duration of a few days. 
With their dynamic range of about 10$^3$-10$^5$, SFXTs are among
the most extreme examples of variability in the X--ray/soft gamma-ray sky.
Their apparently short duty cycle as well as transitory nature are the main reasons
why they escaped detection in the previous  forty  years of X--ray
observations.
In fact, since SFXTs  occur at unpredictable locations  and times, it is very difficult
to detect and discover them using traditional narrow field X-ray  
instruments.
On the contrary, detectors having a sufficiently wide field of view, 
such as IBIS,  are particularly
suited to detect SFXTs, allowing
a greater chance of serendipitously detecting a short duration X--ray event.

The physical mechanism driving the peculiar transient X--ray emission 
from SFXTs is unclear and still highly
 debated: some models involve the structure of the supergiant wind 
(likely clumpy, in a spherical 
 or non spherical geometry; in't Zand 2005; Negueruela et al. 2006; 
Sidoli et al. 2007; Ducci et al. 2009) 
and the orbital properties (wide separation with eccentric or circular 
 orbit), while others involve the properties of the neutron star and 
invoke very low 
 magnetic field values (B$<$10$^{10}$~G; Sguera et al. 2009a) 
or alternatively very high (B$>$10$^{14}$~G, magnetars; Bozzo et al. 2008).
 The picture is still highly unclear from the observational point of view: no cyclotron lines 
 have been detected to date in the SFXTs spectra, thus the strength of the 
neutron star magnetic
 field is unknown. Only in the SFXT IGR J18483--0311 a hint of a 
cyclotron emission line (which however needs confirmation) at 3.3 keV has 
been found in the \xmm\ spectrum, translating into a magnetic field of a 
few B$\sim$10$^{11}$~G (Sguera et al. 2009b). 

 Periodicities, likely orbital, have been measured in five systems, 
spanning from only 3.3 days
 (IGRJ16479--4514; Jain et al. 2009) to 165 days 
(in IGRJ11215--5952; Sidoli et al. 2006, 2007; Romano et al. 2009a). 
 Especially the shortest orbital period is puzzling, because it is even 
shorter than what is observed
in several persistent HMXBs. Indeed, the link between HMXBs displaying
 persistent X--ray emission and SFXTs, is still unclear, and needs an in-depth investigation.

 Even the SFXTs duty cycle seems to be quite different
 among the members of the class (Romano et al. 2009b). 
 Thus a unified picture of the actual mechanism driving the outburst
 emission is still lacking, also in comparison with persistent HMXBs hosting a similar donor star.

\section{SFXTs science with EXIST}

The Energetic X--ray Imaging Survey Telescope ($EXIST$) will be
particularly suited to the detection and subsequent study of new or already known 
Supergiant Fast X--ray Transient sources, since it will provide a very powerful 
combination of very large field of view (FOV), sensitivity  and angular resolution, 
wide energy range coverage and accurate source positioning.

$EXIST$, with its hard X-ray imaging (5--600 keV) 
 all-sky deep survey and large improved  limiting sensitivity, will allow to explore the transient 
 X--ray Universe to get a clearer picture of this new class of puzzling X--ray transients. 
 A complete census of the number of fast transients is essential to enlarge the sample and possibly
 discover different behaviors and SFXTs sub-classes. 
 A long term and continuous as possible X--ray 
 monitoring of SFXTs is crucial to (1)-properly obtain the duty cycle, (2)-to investigate their orbital
 properties (separation, orbital period, eccentricity) which are unknown for the great majority of 
 SFXTs  and search for periodicities predicted by models, (3)-to completely cover the whole outburst 
 activity (now possible only for SFXTs with periodically recurrent X-ray flaring activity, see Figure 1), (4)-to search
 for cyclotron lines in the high energy spectra which could provide the first firm measurement of the 
 neutron star magnetic field. EXIST observations will help in shedding light also to the link between 
 SFXTs and classical HMXBs which are persistent X-ray emitters.
We estimate that EXIST, given its sensitivity about at least one order
of magnitude higher than IBIS, will discover hundreds new SFXTs
on the Galactic plane during one year scanning.

\begin{figure}
  \includegraphics[height=.69\textheight,angle=-90]{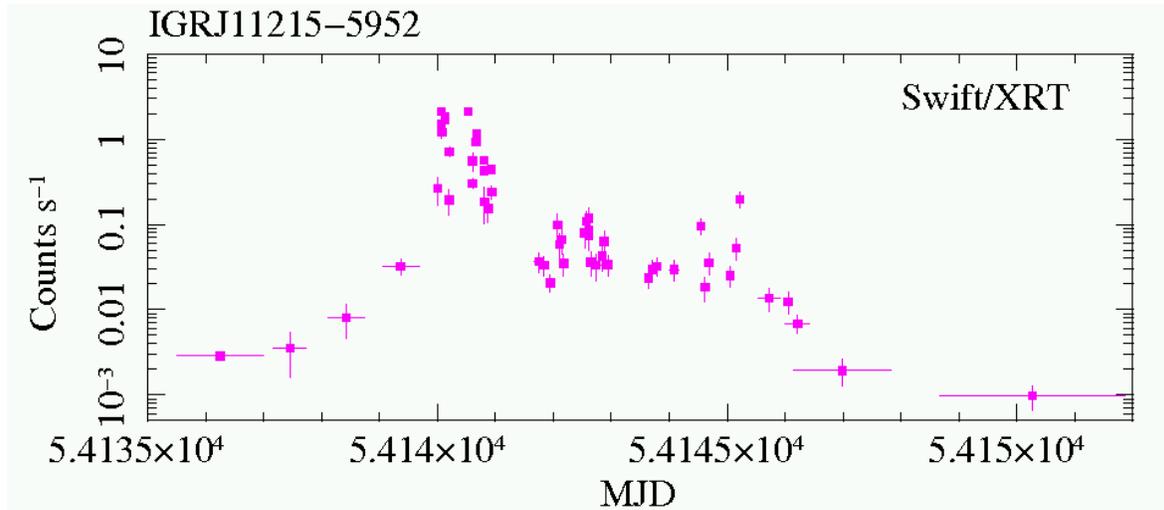}
  \caption{$Swift$/XRT lightcurve of the SFXT IGRJ11215--5952, during one of its periodic outbursts,
in 2007 February}
\end{figure}


\begin{thebibliography}{99}
 
\bibitem{} Bozzo, E., Falanga, M., Stella, L., 2008, \emph{ApJ}, {\bf 683}, 1031 \\
\bibitem{} in't Zand, J.J.M., 2005, \emph{A\&A}, {\bf 441}, L1 \\
\bibitem{} Jain, C, Paul, B, Dutta, A., 2009, \emph{MNRAS}, {\bf 397}, L11 \\
\bibitem{} Negueruela, I., Smith, D.M., Reig, P., et al. 2006, in \emph{ESA Sp. Pub.}, 
ed. A.Wilson, Vol. {\bf 604}, 165-170 \\
\bibitem{} Romano, P., Sidoli, L., Cusumano, G., et al.,  2009a, \emph{ApJ}, 696, 2068 \\
\bibitem{} Romano, P., Sidoli, L., Cusumano, G., et al.,  2009b, \emph{MNRAS}, 399, 2021 \\
\bibitem{} Sguera, V., Barlow, E.J., Bird, A.J., et al. 2005, \emph{A\&A},  {\bf  444}, 221 \\
\bibitem{} Sguera, V., Romero, G. E., Bazzano, A., et al. 2009, \emph{ApJ},  {\bf  697}, 1194 \\
\bibitem{} Sguera, V., Ducci, L., Sidoli, L.,  Bazzano, A., Bassani, L., 2009, \emph{MNRAS} in press, (arXiv 0912.1730) \\
\bibitem{} Sidoli, L., Paizis, A., \& Mereghetti, S., 2006, \emph{A\&A}, 450, L9 \\
\bibitem{} Sidoli, L., Romano, P., Mereghetti, S., et al.,  2007, \emph{A\&A}, 476, 1307 \\



\end{thebibliography}
\end{document}